\documentclass[preprint,preprintnumbers,amsmath,amssymb,floatfix,nofootinbib]{revtex4}
\usepackage{epsf}
\usepackage{graphicx}

\begin{document}

\title{\vspace*{0.7in}
A Higgs Conundrum with Vector Fermions}
\author{S.~Dawson and E.~Furlan}

\affiliation{
Department of Physics, Brookhaven National Laboratory, 
Upton, NY 11973, USA
\vspace*{.5in}}

\begin{abstract}
Many models of Beyond the Standard Model physics involve heavy
colored
fermions. We study models where the new fermions have
vector interactions and examine the connection between 
 electroweak precision measurements and
Higgs production. In particular, for parameters which are allowed
by precision measurements, we show that the gluon fusion Higgs cross section 
and the Higgs decay branching ratios
must be close to those predicted by the Standard Model. The models we discuss
thus represent scenarios with new physics which will be extremely difficult
to distinguish from the minimal Standard Model.
 We pay particular attention to the
decoupling properties of the vector fermions.
\end{abstract}

\maketitle
\newpage

\section{Introduction}

The Standard Model of particle physics has a remarkable body of experimental support, but the Higgs boson
remains a missing ingredient. Precision electroweak measurements suggest that a Standard Model Higgs boson
must be lighter than $\sim 145$~GeV~\cite{Arbuzov2006728, Flacher:2008zq} and recent measurements from the LHC 
exclude a Standard Model Higgs boson in the range $129{\rm ~GeV} < M_H < 600$~GeV~\cite{Chatrchyan:2012tx}. Preliminary measurements
suggest a light Higgs boson in the mass region
$M_H\sim 125$~GeV~\cite{ATLAS:2012ae,Chatrchyan:2012tx}. Should this putative Higgs signal be confirmed, the pressing issue will be understanding its
properties.

For all Higgs masses, gluon fusion is the dominant production mechanism at hadron colliders and the production rate is well
understood up to NNLO in QCD~\cite{Anastasiou:2004xq,Harlander:2002wh}. 
Theoretical uncertainties from renormalization/factorization scale choices and from 
the choice of parton distribution functions
are also well understood~\cite{LHCHiggsCrossSectionWorkingGroup:2011ti,Anastasiou:2011pi,deFlorian:2009hc,Anastasiou:2008tj}.
The total rate, however, is sensitive to the existence of colored particles which couple to
the Higgs boson. Beyond the Standard Model physics can potentially
 have a large effect on the Higgs boson production rate, 
making this a window to high scale 
physics~\cite{Anastasiou:2010bt,Anastasiou:2011qw,Furlan:2011uq,Anastasiou:2011pi}.

The effects on Higgs production of squarks, Kaluza Klein colored fermions, color octet scalars, fermionic
 top quark partners and $4^{\rm th}$ generation fermions (among
 many others) have been extensively studied.
 The simplest possibility for new heavy fermions is to form a chiral
 heavy new generation which, except for masses, 
 is an exact copy of the known generations. After
careful tuning, it is possible to find combinations of $4^{\rm th}$ generation fermion masses
which are permitted by precision electroweak measurements~\cite{Kribs:2007nz,Eberhardt:2010bm}
and are not excluded by direct searches.
Since a chiral  $4^{\rm th}$ generation quark is assumed 
to couple to the Higgs boson with a strength proportional to its mass, heavy
quarks do not decouple from the production of the Higgs boson (and in
fact increase the rate by a factor of $\sim 9$). 
the existence of a $4^{\rm th}$ generation of fermions would exclude a 
Higgs boson mass up to $M_H\sim 600$~GeV~\cite{SM4-LHC} regardless of the fermion masses.

In this paper, we study the effect of heavy vector quarks on Higgs boson production 
and study both the case of an isospin singlet top partner and an isospin doublet of heavy fermions. 
A vector singlet top partner arises naturally in Little Higgs 
models~\cite{ArkaniHamed:2002qy,Low:2002ws,Perelstein:2003wd,Chang:2003un,Chen:2003fm,Hubisz:2005tx,Han:2005ru}, 
where the couplings to the Higgs boson of the top
quark and its fermion partner are fixed in such a manner as to cancel their
quadratically divergent contributions to the Higgs mass renormalization. 
Top-quark fermion partners are also found
in top color~\cite{Hill:1991at,Hill:2002ap} and top condensate~\cite{Dobrescu:1997nm,Chivukula:1998wd,He:1999vp,Fukano:2012qx}
models where there is a natural hierarchy of scales such that
the top partner obtains a large Dirac mass. 
Light vector fermions instead typically appear in composite Higgs 
models~\cite{Contino:2006qr,Carena:2006bn,Barbieri:2007bh,Matsedonskyi:2012ym}. 
Our results are general enough to be applied to any of these
models and hence represent a simplification of results which have previously been presented in
the context of very specific scenarios. 

A study of the $S$, $T$, $U$ parameters 
and the $Z\rightarrow
b{\overline b}$ decay rate~\cite{Bamert:1996px,Comelli:1996xu} restricts the
allowed parameter space for heavy vector fermions. However, vector fermions have interesting decoupling
properties as the mixing with the
Standard Model fermions becomes small, which makes
a large region of parameter space experimentally viable.
Vector fermions which couple to the Standard Model fermions and Higgs boson can be
 $SU(2)_L$ singlets with the same hypercharge as the Standard Model right-handed quarks,
 doublets with $Y=Y_{SM}={1\over 6}$ or $Y=Y_{SM}\pm 1$,  or triplets with $Y=Y_{SM}\pm{1\over 2}$~\cite{Cacciapaglia:2010vn}.
We consider the ``Standard Model-like" case with either a heavy fermion singlet of charge 2/3 or a doublet with 
the Standard Model assignments of hypercharge.
We compute the NNLO prediction for Higgs production for the allowed
parameter region of 
these models and quantify the allowed deviation from the Standard Model prediction.
The new features of our study include up-to-date
fits to precision electroweak measurements in models with vector
fermions, and an analysis of the the resulting
consequences for Higgs boson production at NNLO in perturbative QCD.

\section{The Models}
We consider models with additional vector-like charge 2/3 quarks, ${\cal{T}}^\alpha$,
and charge -1/3 quarks, ${\cal{B}}^\alpha$, which mix with the Standard Model-like 
third generation quarks. For simplicity we make the following assumptions:
\begin{itemize}
\item the electroweak gauge group is the standard $SU(2)_L\times U(1)_Y$ group;
\item there is only a single Standard Model Higgs $SU(2)_L$ doublet,
\begin{equation}
H = \left(\begin{matrix}\phi^+\\ \phi^0\end{matrix}\right) \,,
\end{equation}
with $\phi^0={v+h \over \sqrt{2}}$;
\item we neglect generalized CKM mixing and only allow mixing 
between the Standard Model-like third generation quarks and at most
one new charge 2/3 quark singlet
or one new $SU(2)_L$ quark doublet. We do not consider fermions in more exotic representations.
\end{itemize}
The Standard Model-like chiral fermions are
\begin{equation}
\psi^1_L=\left(\begin{matrix}
{\cal{T}}_L^1\\
{\cal {B}}_L^1\end{matrix}\right), \quad {\cal{T}}^1_R, {\cal{B}}^1_R\, ,
\end{equation}
with the Lagrangian describing the fermion masses 
\begin{equation}
-{\cal L}_M^{SM}=\lambda_1{\overline{\psi}}^1_L H {\cal {B}}^1_R+
\lambda_2{\overline{\psi}}^1_L {\tilde H} {\cal{T}}^1_R+{\rm h.c.} \, ,
\end{equation}
and ${\tilde H}=i\sigma_2 H^*$.

The models we consider are:
\begin{itemize}
\item singlet fermion model: add a vector $SU(2)_L$ quark singlet
of charge 2/3, ${\cal {T}}^2_L$ and ${\cal {T}}^2_R$.

\item doublet fermion model:
add a vector $SU(2)_L$ doublet of hypercharge 1/6,
\begin{equation}
\psi^2_L=
\left( 
\begin{matrix} 
{\cal {T}}_L^2\\
{\cal {B}}_L^2
\end{matrix}\right),
\quad
\psi^2_R=
\left(
\begin{matrix}
{\cal {T}}^2_R\\
{\cal {B}}^2_R\end{matrix}
\right) \, .
\end{equation}

\end{itemize}

\section{Experimental Limits on Top Partner Models}
\label{exp}
\subsection{Limits from $R_b$ and $A_b$}
\label{bfits}

Data from LEP and SLD 
place stringent restrictions on the couplings of the fermionic top partners.
The top partners mix with the Standard Model-like top quark and contribute at one loop to processes
involving bottom quarks, especially $Z\rightarrow b {\overline b}$ and $A_b$. The neutral current couplings to the 
bottom can be parametrized by the effective Lagrangian 
\begin{equation}
{\cal L}^{NC}={g\over c_W}Z_\mu {\overline f}\gamma^\mu
\biggl[\biggl(g_L^f+\delta{\tilde g}_L^f\biggr)
\biggl({1-\gamma_5\over 2}\biggr)
+\biggl(g_R^f+\delta{\tilde g}_R^f\biggr)\biggl({1+\gamma_5\over 2}
\biggr)\biggr] f \, ,
\label{zdef}
\end{equation}
where the Standard Model couplings are normalized such that 
$g_L^f=T_3^f - Q_f s_W^2$, 
\mbox{$g_R^f=-Q_f s_W^2$}, with $s_W^2\equiv \sin^2\theta_W=(e/g)^2=0.231$~\cite{Nakamura:2010zzi} and
$T_3^f=\pm 1/2$. 
The couplings 
$\delta {\tilde g}_{L,R}^f
\equiv\delta g_{L,R}^{f,SM}+\delta g_{L,R}^f$ contain both the Standard
Model radiative corrections, $\delta g_{L,R}^{f,SM}$, 
and the new physics contributions, $\delta g_{L,R}^{f}$.
The Standard Model contribution from top quark 
loops is well known~\cite{Akhundov:1985fc,Bernabeu:1987me,Beenakker:1988pv}, 
and in the limit $m_t>>M_Z$ it is given by
\begin{equation}
\delta g_L^{b,SM} = {G_F\over\sqrt{2}}{m_t^2\over 8\pi^2}\, .
\end{equation}

The dominant effect of new physics in the $b$ sector 
can be found by assuming that $\delta g_L^b$ and $\delta g_R^b$ are 
small and approximating~\cite{Burgess:1993vc, Haber:1999zh}
\begin{eqnarray}
R_b&=& {\Gamma(Z\rightarrow b {\overline b})
\over \Gamma(Z\rightarrow hadrons)}
=R_b^{SM}\biggl\{1-3.57\delta g_L^b+0.65\delta g_R^b\biggr\}\nonumber \\
A_b&=&{(\delta \tilde{g}_L^{b})^2
-(\delta \tilde{g}_R^{b})^2 \over
(\delta \tilde{g}_L^{b})^2+
(\delta \tilde{g}_R^{b})^2}=A_b^{SM}\biggl\{1-0.31\delta g_L^b-1.72 \delta g_R^b\biggr\} \,,
\end{eqnarray}
where $R_b^{SM}$ and $A_b^{SM}$ are
the theory predictions including all radiative corrections. 
The positive contribution to $\delta g_L^{b,SM}$ from the top quark has the effect of 
reducing both $R_b^{SM}$ and $A_b^{SM}$. 

\begin{figure}
\begin{center}
\includegraphics[bb=14 38 509 390, scale=0.6]{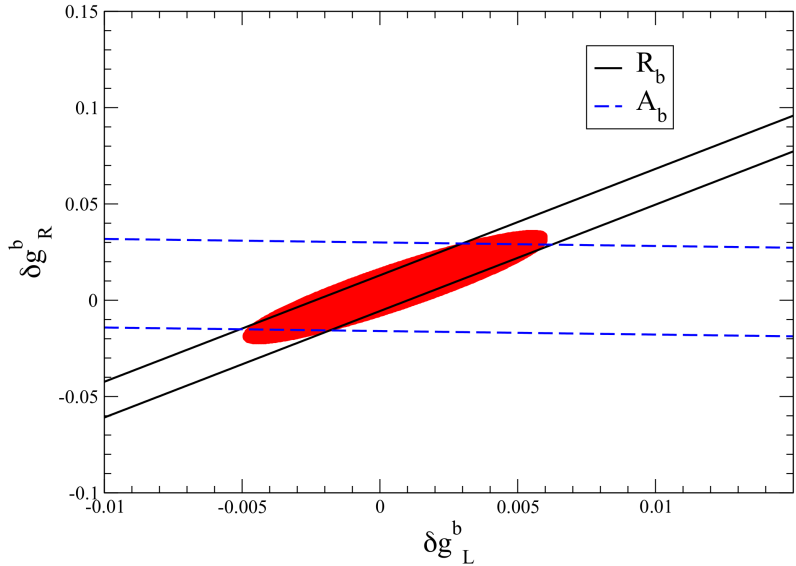} 
\vskip .25in
\caption[]{Allowed $95\%$ confidence level regions from the simultaneous 
fit to $R_b$ and $A_b$ (red shaded),
$R_b$ alone 
(between solid black lines), and $A_b$ alone (between dashed blue
lines).}
\label{fg:zbb}
\end{center}
\end{figure}
The $95\%$ confidence level ellipse for new $Zb{\overline b}$ couplings is shown in 
Fig.~\ref{fg:zbb} and is obtained 
using the Particle Data Group results~\cite{ALEPH:2005ab, Nakamura:2010zzi} 
\begin{eqnarray}
R_b^{exp}&=& 0.21629\pm 0.00066 \nonumber \\
R_b^{SM}&=& 0.21578\pm 0.00005\nonumber \\
A_b^{exp}&=&0.923\pm 0.020\nonumber \\
A_b^{SM}&=&0.9348\pm 0.0001 \, .
\end{eqnarray}
If $\delta g_R^b=0$, the $95\%$ confidence level limit from the
fit to $A_b$ and $R_b$ is 
\begin{equation}
-0.0027 < \delta g_L^b < 0.0014 \, .
\end{equation}
Similarly, if $\delta g_L^b=0$, the $95\%$ confidence level limit from the
fit to $A_b$ and $R_b$ is
\begin{equation}
-0.0066 < \delta g_R^b < 0.0148 \, .
\label{grfit}
\end{equation}

\subsection{Limits from the Oblique Parameters $S$, $T$ and $U$}
\label{stufits}
The new quarks contribute at loop level to the vacuum polarizations 
of the electroweak gauge bosons 
$\Pi_{XY}^{\mu\nu}(p^2)=\Pi_{XY}(p^2)g^{\mu\nu}+B_{XY}(p^2)p^\mu p^\nu$,
with $XY=\gamma\gamma, \gamma Z, ZZ$ and $W^+W^-$~\cite{Peskin:1991sw,Altarelli:1990zd}. 
These effects can be parametrized using the $S, T$ and $U$ functions
of Peskin and Takeuchi~\cite{Peskin:1991sw},
\begin{eqnarray}
\alpha S_F &=&
{4 s_W^2 c_W^2\over M_Z^2}
\biggl\{ 
\Pi_{ZZ}(M_Z^2)- \Pi_{ZZ}(0)-\Pi_{\gamma\gamma}(M_Z^2)
-{c_W^2-s_W^2\over c_W s_W}
\Pi_{\gamma Z}(M_Z^2)
\biggr\}
\nonumber \\
\alpha T_F &=&
{ \Pi_{WW}(0)\over M_W^2}
-{\Pi_{ZZ}(0)\over M_Z^2}
\nonumber \\
\alpha U_F&=& 4 s_W^2\biggl\{
{ \Pi_{WW}(M_W^2)-\Pi_{WW}(0)\over M_W^2} 
-c_W^2\biggl({ \Pi_{ZZ}(M_Z^2)-\Pi_{ZZ}(0)\over M_Z^2}\biggr)
\nonumber \\
&&-2 s_W c_W
{ \Pi_{\gamma Z}(M_Z^2)
\over M_Z^2}
-s_W^2 { \Pi_{\gamma \gamma}(M_Z^2)\over M_Z^2}\biggr\} \, .
\label{sdef}
\end{eqnarray}
Any definition of $s_W$ can be used in Eq.~\ref{sdef} since
the scheme dependence enters at higher order. Since these 
parameters are well-constrained by LEP and LEP2 measurements~\cite{Barbieri:2004qk}, 
they set stringent limits on the masses and couplings of the new quarks. 

We use the fit to the electroweak
precision data given in Refs.~\cite{Flacher:2008zq,Baak:2011ze}, 
\begin{eqnarray}
\Delta S = S-S_{SM}&=& 0.02 \pm 0.11 \nonumber \\
\Delta T = T-T_{SM}&=&0.05\pm 0.12\nonumber \\
\Delta U = U-U_{SM}&=&0.07 \pm 0.12 \, ,
\label{delts}
\end{eqnarray}
with reference Higgs and top-quark masses  
$M_{H,{\rm ref}}=120$~GeV and $m_{t,{\rm ref}} = 173.1$~GeV. 
The associated correlation matrix is 
\begin{eqnarray}
\rho_{ij}=\left(
\begin{array}{lll}
1.0 & 0.879 & -0.469\nonumber \\
0.879 & 1.0 & -0.716 \nonumber\\
-0.469 & -0.716 & 1.0
 \end{array}
\right)\, .
\end{eqnarray}
The $\Delta \chi^2$ is defined as
\begin{equation}
\Delta \chi^2=\sum_{i,j}(\Delta X_i-\Delta {\hat X}_i)
(\sigma^2)^{-1}_{ij}(\Delta X_j-\Delta {\hat X}_j)\, ,
\end{equation}
where $\Delta {\hat X}_i =\Delta S, \Delta T$ and
$\Delta U$ are the central values of the fit in Eq.~\ref{delts}, 
$\Delta X_i=X_i-X_i^{SM}=\Delta S_F, \Delta T_F$ and
$\Delta U_F$ are the contributions to the oblique parameters from the new 
fermions and $\sigma^2_{ij}\equiv \sigma_i\rho_{ij}\sigma_j\,$, $\, \sigma_i$ being the errors
given in Eq.~\ref{delts}.
A $95\%$ confidence level limit in a three-parameter fit corresponds to $\Delta \chi^2 = 7.82$.

Since we consider primarily $M_H=125$~GeV, we need to add the Higgs 
contributions\footnote{Our fits include the exact results for the Higgs contributions, which can
be found in many places including Ref.~\cite{Hollik:1993xg} and the Appendix of Ref.~\cite{Chen:2008jg}.}
\begin{eqnarray}
\Delta S_H&=& {1\over 12 \pi}
\log\biggl(
{M_H^2\over M_{H,{\rm ref}}^2}
\biggr)
+{\cal O}
\biggl(
{M_Z^2\over M_H^2}\biggr)
\nonumber \\
\Delta T_H&=& -{3\over 16 \pi c_W^2}
\log\biggl(
{M_H^2\over M_{H,{\rm ref}}^2}\biggr)
+{\cal O}\biggl({M_Z^2\over M_H^2}
\biggr)
\nonumber \\
\Delta U_H&=& 
{\cal O}\biggl({M_Z^2\over M_H^2}\biggr)\, .
\end{eqnarray}
\subsection{Other Experimental Limits on Top Partner Models}

Both ATLAS~\cite{Aad:2012xc,atlasbs} and CMS~\cite{cmsbs,Collaboration:2012ye} 
have searched for direct pair production of new heavy fermions. For charge 2/3 top-like quarks
decaying with $100\%$ branching ratio to $Wb$, CMS excludes masses below $557$~GeV at $95\%$ confidence
level, while ATLAS sets an upper bound of 404 GeV. CMS dedicates a specific analysis to pair-produced 
vector quarks of charge 2/3 decaying entirely to $Zt$, excluding masses below 475~GeV~\cite{Chatrchyan:2011ay}. 
For charge -1/3 quarks, assuming 
$100\%$ branching ratio to $Zb$, ATLAS excludes masses 
below 358~GeV for a vector singlet, while CMS excludes charge -1/3 quarks decaying with $100\%$ branching ratio 
to $Wt$ below $611$~GeV. These limits are not directly applicable to our models, since the branching ratios 
of the new heavy fermions to Standard Model particles are degraded
by mixing angles and the limits 
therefore weakened~\cite{AguilarSaavedra:2009es,Cacciapaglia:2010vn,Flacco:2011ym,Cacciapaglia:2011fx,Harigaya:2012ir,
Berger:2012ec}. 
Our results on Higgs production are rather insensitive to the masses of the new top
partners and we typically assume masses of the TeV scale. 

In principle, there are also limits on heavy charged fermions which mix with the Standard Model third generation
quarks coming from $K$, $B$ and $D$ rare processes. For TeV-scale masses of the new fermions and
small mixing parameters (which we will see in the next section are required by limits from oblique parameters,
$R_b$ and $A_b$), the constraints from rare processes are not 
restrictive~\cite{AguilarSaavedra:2002kr,Blanke:2006sb,Cacciapaglia:2011fx,Berger:2012ec}.

\section{Singlet Top Partner Model}
Little Higgs models~\cite{ArkaniHamed:2002qy,Low:2002ws,Perelstein:2003wd,
Chang:2003un,Chen:2003fm,Hubisz:2005tx,Han:2005ru}, 
topcolor models~\cite{Hill:1991at,Hill:2002ap} and 
top condensate models~\cite{Dobrescu:1997nm,Chivukula:1998wd,He:1999vp} 
all contain a charge 2/3 partner of the top quark, which we denote
by ${\cal {T}}^2$. We consider a general case
with a vector $SU(2)_L$ singlet fermion which is allowed
to mix with the Standard Model-like top 
quark~\cite{Cacciapaglia:2010vn,Perelstein:2003wd,Lavoura:1992qd,AguilarSaavedra:2002kr,
Popovic:2000dx}.
The mass eigenstates are $b\equiv {\tilde{{\cal{B}}}}^1$, $t$ and $T$,
where $b$ and $t$ are the observed bottom and top quarks. 
Thorough this paper we will use the measured mass values 
$m_b = 4.19$~GeV, \mbox{$m_t = 173.1$~GeV}~\cite{Group:2009ad, Nakamura:2010zzi}. 
The mass eigenstates of charge 2/3 can be found through the rotations
\begin{equation}
\chi_{L,R}^t\equiv
\left( \begin{matrix} 
t_{L,R}\\
T_{L,R}\end{matrix}\right)
\equiv U_{L,R}^t
\left(\begin{matrix}
{\cal{T}}^1_{L,R}\\{\cal{T}}^2_{L,R}
\end{matrix}
\right)\, .
\label{chit}
\end{equation}
The matrices $U_{L,R}^t$ are unitary and $\Psi_{L,R}\equiv{1\mp\gamma_5\over 2}\Psi$.
The mixing matrices are parametrized as
\begin{eqnarray}
U_L^t&=&
\left(\begin{matrix}
\cos\theta_L& -\sin\theta_L\\
\sin\theta_L& \cos\theta_L\end{matrix}
\right),\quad 
U_R^t=
\left(\begin{matrix}
\cos\theta_R & -\sin\theta_R\\
\sin\theta_R & \cos\theta_R\end{matrix}
\right) \, .
\end{eqnarray}
We abbreviate $c_L\equiv \cos \theta_L$, $s_L\equiv \sin\theta_L$.

The most general fermion mass terms are
\begin{eqnarray}
-{\cal L}_{M,1}
&=&
-{\cal L}_M^{SM}+
\lambda_3 
{\overline {\psi}}^1_L
{\tilde {H}} 
{\cal {T}}^2_R+
\lambda_4
{\overline{{\cal {T}}}}^2_L 
{\cal {T}}^1_R+\lambda_5
{\overline{{\cal{T}}}}^2_L
{\cal {T}}^2_R+{\rm h.c.}
\nonumber \\
&=&
{\overline {\chi}}_L^t \biggl[U_L^t \biggl( M^t_{(1)} + h \, H^t_{(1)} \biggr) U_R^{t, \dagger}\biggr]\chi_R^t +
\lambda_1{v+h\over\sqrt{2}} \, {\overline{\cal{B}}}^1_L{\cal {B}}^1_R+ {\rm h.c.}\, ,
\label{all}
\end{eqnarray}
where
\begin{equation}
M^t_{(1)}=\left(
\begin{matrix}\lambda_2{v\over\sqrt{2}}&\lambda_3{v\over\sqrt{2}}\\
\lambda_4&\lambda_5\end{matrix}
\right)
\quad, \qquad
H^t_{(1)}={1 \over \sqrt{2}} \left(
\begin{matrix} \; \lambda_2 \;  & \; \lambda_3 \; \\
0 & 0\end{matrix}
\right)
\, .
\label{eq:M0singlet}
\end{equation}
The resulting mass eigenstates are
\begin{equation}
{\cal M}^{t,diag}\equiv\left(\begin{matrix} m_t & 0 \\
0&M_T\end{matrix} \right)\, .
\end{equation}

One can always rotate ${\cal T}^2$ such that $\lambda_4 = 0$. 
Since $\lambda_4$ can be rotated away, the model has four free parameters. 
Alternatively, it is always possible to rotate ${\cal T}^2_R$ such that $\sin\theta_R=0$, 
because only the Standard Model-like left-handed doublet $ {\psi}^1_L$ mixes to the singlet 
with a Yukawa term\footnote{
	We do not perform any of these rotations here, and the formulas in this section hold for the arbitrary
Yukawa couplings of Eq.~\ref{all}. }. 
Therefore the couplings only depend on $\theta_L$, which we will 
take as one of the four physical parameters along with $m_b$ 
(physical mass of the charge -1/3 quark), $m_t$ and $M_T$ (physical masses of the charge 2/3 quarks).

The physical masses and mixing angles are found using
bi-unitary transformations, 
\begin{eqnarray}
\biggl({\cal M}^{t,diag}\biggr)^2
&=&
U_L^t M_{(1)}^t M_{(1)}^{t,\dagger} U_L^{t,\dagger} 
\nonumber \\
&=&
U_R^t M_{(1)}^{t,\dagger}  M_{(1)}^t U_R^{t,\dagger}
\, . 
\end{eqnarray}
It is straightforward to find the mass eigenstates and mixing angles,
\begin{eqnarray}
\tan (2\theta_R)
&=&
{2\lambda_4\lambda_5 +v^2\lambda_2\lambda_3
\over \lambda_5^2-\lambda_4^2+{v^2\over 2}(\lambda_3^2-\lambda_2^2)}
\nonumber \\
\tan (2\theta_L)
&=&
{\sqrt{2}v(\lambda_2\lambda_4 +\lambda_3\lambda_5)
\over \lambda_5^2+\lambda_4^2-{v^2\over 2}(\lambda_2^2+\lambda_3^2)}
\nonumber \\
m_t M_T
&=&
 { v\over \sqrt{2}}\mid \lambda_2\lambda_5-\lambda_3\lambda_4\mid
 \nonumber \\
M_T^2+m_t^2
&=&
 {v^2\over 2}(\lambda_2^2+\lambda_3^2)+\lambda_4^2+\lambda_5^2\,.
\label{forms}
\end{eqnarray} 
From Eq.~\ref{all}, the couplings to the Higgs boson are 
\begin{equation}
{\cal L}^h_1=-{m_t\over v}c_{tt}{\overline t}_Lt_Rh
-{M_t\over v} c_{TT}{\overline T}_LT_Rh
-{M_t\over v} c_{tT}{\overline t}_L T_Rh
-{m_t\over v} c_{Tt}{\overline T}_Lt_Rh
+{\rm h.c.} \, ,
\end{equation}
where
\begin{equation}
c_{tt}= c_L^2\, , \qquad c_{TT}=s_L^2\, ,\qquad c_{tT}=c_{Tt}=s_Lc_L \, .
\label{yuksing}
\end{equation}
%
These relations can be easily derived by noticing that
\begin{equation}
\label{eq:rel_MH_singlet}
	v H^t_{(1),ks} = M^t_{(1),ks} \delta_{k 1} \, ,
\end{equation}
yielding for the physical Higgs couplings
\begin{eqnarray}
\label{eq:why_UR_vanishes_singlet}
	H_{ij}& \equiv & U_{L,ik}^t H^t_{(1),ks} U_{R,sj}^{t,\dagger}
	\nonumber \\
	& = &
	v^{-1} U_{L,i \hat{k}}^t \delta_{\hat{k} 1} \biggl[
		U_{L, \hat{k} r }^{t,\dagger} {\cal M}^{t,diag}_{rr} U_{R,rs}^t
	\biggr]
	U_{R,sj}^{t,\dagger}
		\nonumber \\
	& = &
	v^{-1} U_{L,i \hat{k}}^t \delta_{\hat{k} 1} U_{L, \hat{k} j }^{t,\dagger} {\cal M}^{t,diag}_{jj} \,,
\end{eqnarray}
where the index $\hat{k}$ is not summed over.

The charged and neutral current interactions are
\begin{eqnarray}
{\cal L}^{CC}_1
&=&
{g\over \sqrt{2}}  W^{\mu +}\sum_{i=1,2}
\biggl({\overline \chi}^t_L\biggr)_{i} \biggl(U_L^t\biggr)_{i,1} \gamma_\mu  b_L + {\rm h.c.} \, 
\end{eqnarray}
and
\begin{eqnarray}
{\cal L}^{NC}_1
&=&
{g\over c_W} Z_\mu
\sum_{i=t,T}
\biggl\{\overline{f_i}\gamma^\mu
\biggl[(g_L^i+\delta {\tilde{g}}_L^i)P_L
+
(g_R^i+\delta {\tilde{g}}_R^i) P_R \biggr]
f_i
\biggr\}
\nonumber \\
&&+
{g\over c_W} Z_\mu
\sum_{i\ne j}
\biggl\{\overline{f_i}\gamma^\mu
\biggl[\delta g_L^{ij} P_L 
+\delta g_R^{ij} P_R \biggr]f_j\biggr\} \, ,
\label{neutc}
\end{eqnarray}
where $\delta{\tilde g}^i_{L,R}=\delta g^{i,SM}_{L,R}+\delta g^i_{L,R}$
contains both the Standard Model contribution from top quark loops and the new physics contributions.
The new physics couplings arising from the interactions of the top partner singlet are
\begin{eqnarray}
\delta g_L^t&=&-{s_L^2\over 2} \quad , \qquad 
\delta g_L^T=-{c_L^2\over 2} \quad , \qquad 
\delta g _L^{tT}={s_Lc_L\over 2} \nonumber \\
\delta g_R^i&=&\delta g_R^{ij}=0 \;\;\,,\quad i,j = t,T \, .
\end{eqnarray}

Finally, the unitarity bound from the scattering $T{\overline T}\rightarrow T{\overline T}$ is
modified from the Standard Model limit and becomes~\cite{Chanowitz:1978mv, Dawson:2010jx}
\begin{equation}
M_T({\hbox{Unitarity~Bound}}) \lesssim {550~{\rm GeV}\over s_L^2}\, .
\end{equation}
This class of models therefore allows quite heavy $T$ quarks without violating perturbative unitarity.

\subsection{Experimental Restrictions on Singlet Top Partner Model}
\label{sec:exp_bounds_singlet}
Using the results given above, it is straightforward to compute the 
contributions to the
$S, T$ and $U$ parameters in the singlet top partner model.
Expressions for the gauge boson two-point functions for arbitrary fermion couplings 
are given in the appendix~\cite{Chen:2003fm,Jegerlehner:1991dq,Cynolter:2008ea},
and we assume $M_T >> M_Z$. 
Subtracting the Standard Model $t-b$ loops, the new contributions are 
\begin{eqnarray}
\Delta T_F&=&
T_{SM} \, s_L^2
\biggl[
-(1+c_L^2) + s_L^2 r + 2 c_L^2{r \over r- 1} \log(r)
+{\cal O}\biggl({M_Z^2\over
m_t^2},{M_Z^2\over M_T^2}, 
{m_b^2\over m_t^2}\biggr)
\biggr]
\nonumber \\
\Delta S_F &=& -{N_C\over 18\pi}s_L^2\biggl\{
\log(r)
+c_L^2
\biggl[
{5(r^2+1)-22r\over (r-1)^2}
+{3(r+1)(r^2-4r+1)\over (1-r)^3}\log(r)
\biggr]
\nonumber \\ 
&& +{\cal O}\biggl({M_Z^2\over
m_t^2},{M_Z^2\over M_T^2}, {m_b^2\over m_t^2}\biggr)
\biggr\}
\nonumber \\
\Delta S_F+ \Delta U_F&=&
{N_C\over 9 \pi}s_L^2 
\biggl[
	\log(r) + {\cal O}\biggl({M_Z^2\over m_t^2},{M_Z^2\over M_T^2}, {m_b^2\over m_t^2}\biggr)
\biggr]\, ,
\label{stusing}
\end{eqnarray}
where
\begin{equation}
r \equiv {M_T^2\over m_t^2} \quad , \quad N_C=3 \;\;\; \textrm{, \, and} \quad 
T_{SM} = { N_C\over 16 \pi s_W^2} {m_t^2 \over M_W^2}\, 
\end{equation}
is the $t - b$ contribution to the $T$ parameter in the limit of a massless bottom quark. 
We use $M_W = 80.4$~GeV, $M_Z = 91.2$~GeV~\cite{Nakamura:2010zzi}. 
Eq.~\ref{stusing} agrees with the $m_b\rightarrow 0$ and $M_Z<<M_T,m_t$ limits of
Refs.~\cite{Maekawa:1995ha,He:2001fz,Csaki:2003si}.\footnote{
There is a typo in Eq.~88 of Ref.~\cite{He:2001fz}, where there is a 2 instead of the factor 22 in Eq.~\ref{stusing}.
}
For a top parter much heavier than the top quark, $M_T>>m_t$, the oblique parameters take simple forms,
\begin{eqnarray}
\Delta T_F({\hbox{approx}}) &= &
T_{SM} \, s_L^2
 \biggl[-(1+c_L^2)+  s_L^2 r + 2 c_L^2\log(r) \biggr]
\nonumber \\
\Delta S_F({\hbox{approx}})&= &
-{N_C\over 18 \pi}s_L^2
\biggl[ 5 \, c_L^2 + (1- 3c_L^2)\log(r)\biggr]\, .
\label{singapprox}
\end{eqnarray}

One would expect decoupling to occur for a very heavy vector top partner, i.e. for 
$r \to \infty$. From Eqs.~\ref{stusing},~\ref{singapprox}, instead, 
the effects of the top partner on the oblique parameters vanish only in the limit $s_L\rightarrow 0$. 
This can be understood inspecting the mass matrix~(\ref{eq:M0singlet}): 
to obtain decoupling both the Yukawa interactions and the off-diagonal terms need to be small, 
$\lambda_2 v,\lambda_3 v,\lambda_4 << \lambda_5$. In this limit  
\begin{equation}
s_L\rightarrow {v\lambda_3 \over\sqrt{2}M_T}+{\lambda_2\lambda_4 v\over \sqrt{2}M_T^2}+...
\end{equation}
and the top partner effects vanish for large $M_T$, as expected. 

\begin{figure}
\begin{center}
\includegraphics[bb=14 38 509 390,scale=0.6]{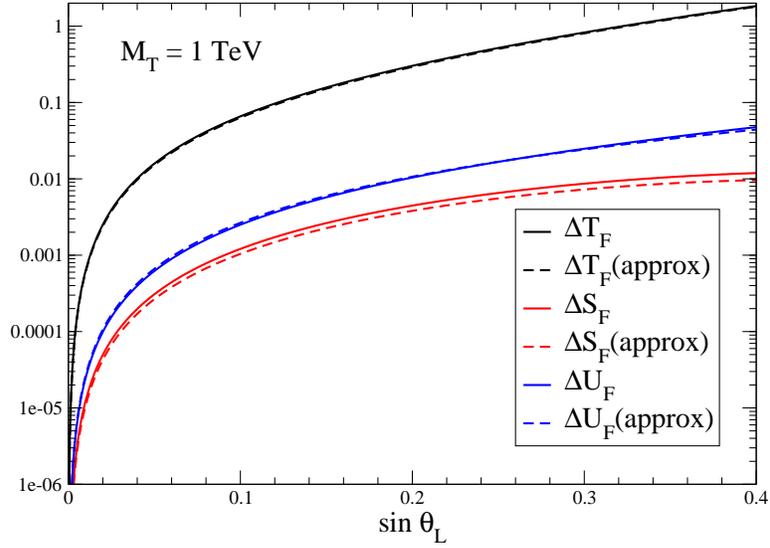}
\caption[]{Contributions to $\Delta T_F$ , $\Delta S_F$, 
$\Delta U_F$ from a 
singlet top partner as a function of $\sin \theta_L$ 
for fixed \mbox{$M_T=1$}~TeV. The results of Eq.~\ref{singapprox} in the
limit $M_T>>m_t$ are shown as $\Delta T_F$(approx), $\Delta S_F$(approx) 
and $\Delta U_F$(approx).
 }
\label{fg:dt}
\end{center}
\end{figure}
In Fig.~\ref{fg:dt} we show the oblique parameters for a fixed $M_T=1$~TeV. It is clear
that the approximate forms of Eq.~\ref{singapprox} are excellent approximations to the complete
results for mass values of this order. 
\begin{figure}
\begin{center}
\includegraphics[bb=14 38 509 390,scale=0.6]{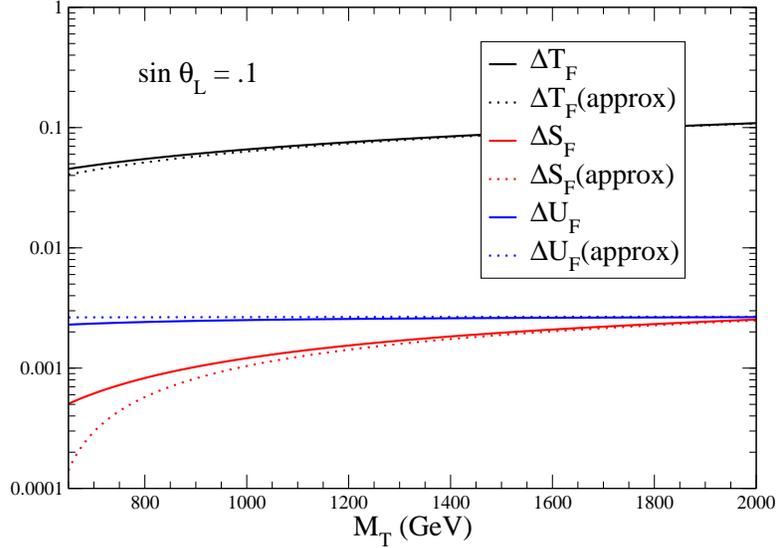} 
\caption[]{
Fermion contributions to $\Delta T_F$ , $\Delta S_F$,
$\Delta U_F$ in the 
singlet top partner model for fixed $\sin\theta_L=0.1$. The dotted lines represent the 
approximate results from Eq.~\ref{singapprox} in the
limit \mbox{$M_T>>m_t$}.
}
\label{fg:dt2}
\end{center}
\end{figure}
The largest contribution is to $\Delta T_F$ due to the 
large isospin violation for non-zero $\sin \theta_L$. In this case, the isospin violation is manifest in the
result that $\Delta U_F>\Delta S_F$. The new physics effects vanish as $s_L\rightarrow 0$
and we recover the Standard Model couplings.
The oblique parameters for fixed
$\sin\theta_L$ are shown in Figs.~\ref{fg:dt2} and~\ref{fg:dt3} as a function of $M_T$. 
As we argued, the decoupling does not occur for large $M_T$, but requires $s_L\rightarrow 0$. 

\begin{figure}
\begin{center}
\includegraphics[bb=14 38 509 390,scale=0.6]{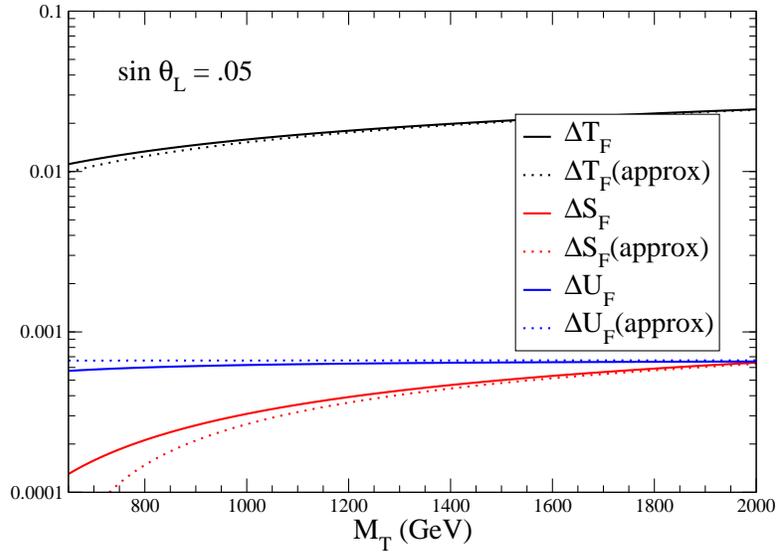} 
\vskip .25in 
\caption[]{
Same as Fig.~\ref{fg:dt2} for a smaller $\sin\theta_L=0.05$. 
 }
\label{fg:dt3}
\end{center}
\end{figure}

\begin{figure}
\begin{center}
\includegraphics[bb=14 38 509 390,scale=0.6]{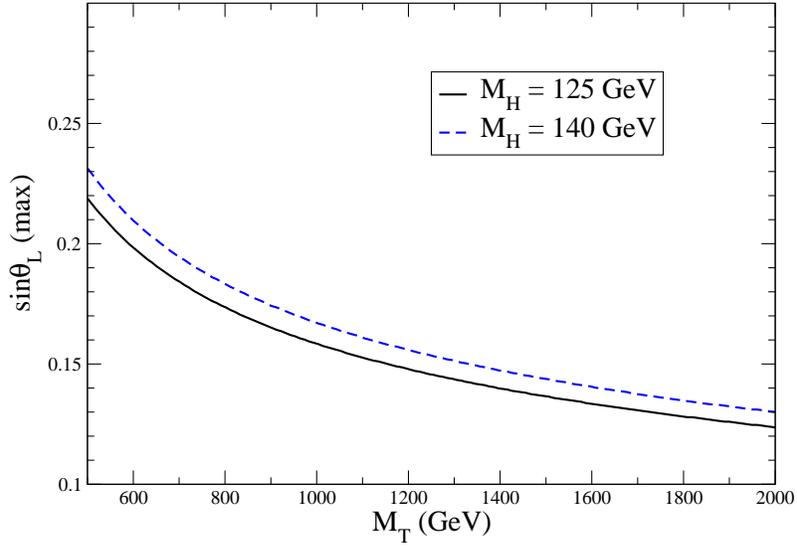} 
\vskip .25in 
\caption[]{$95\%$ confidence level upper bound on
 the mixing angle $\sin\theta_L$ in the singlet
top partner model from experimental restrictions on the $S$, $T$ and $U$ parameters.}
\label{fg:stulim1}
\end{center}
\end{figure}

\begin{figure}
\begin{center}
\includegraphics[bb=14 38 509 390,scale=0.6]{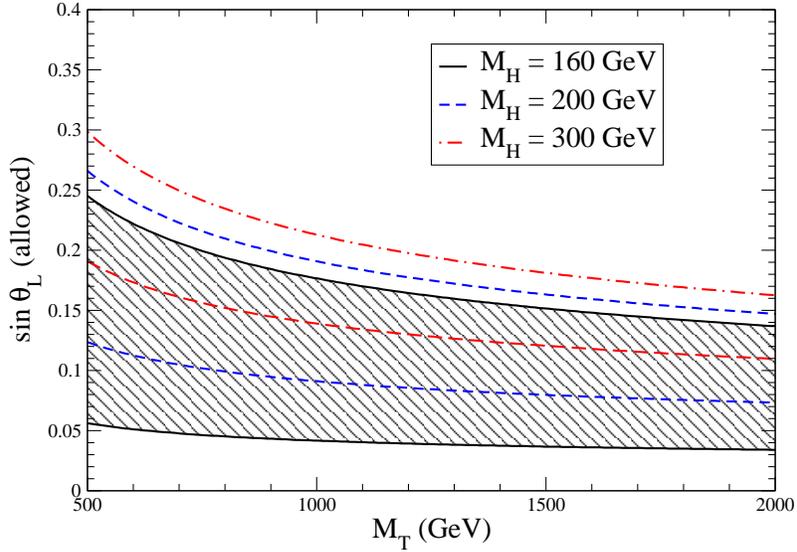} 
\vskip .25in 
\caption[]{$95\%$ confidence level bands allowed by a fit to $S$, $T$ and $U$ for the mixing angle 
$\sin\theta_L$ in the singlet
top partner model. The regions allowed are between the curves corresponding to
each value of $M_H$.}
\label{fg:stulim2}
\end{center}
\end{figure}

The parameter space allowed by a fit to the oblique parameters can be found using the results of 
Section~\ref{exp}. 
Fig.~\ref{fg:stulim1} shows the $95\%$ confidence level upper bound 
on the mixing
angle $\sin\theta_L$ as a function of $M_T$ for a light Higgs boson. 
For a heavier Higgs boson, it is possible
to use the positive contribution to $T$ from the top partner to compensate for the negative contribution from the heavy Higgs,
as shown in Fig.~\ref{fg:stulim2}. 
In this case, a minimum degree of mixing is required, since such a heavy Higgs boson is excluded by 
the electroweak fit in the three-generation Standard Model.
The heavier the Higgs boson, the smaller
the range of $\sin\theta_L$ allowed. This situation was explored for an extremely heavy Higgs boson
($M_H\sim800$~GeV) in Ref.~\cite{Bai:2011aa}.

\begin{figure}
\begin{center}
\includegraphics[bb=14 38 509 390,scale=0.6]{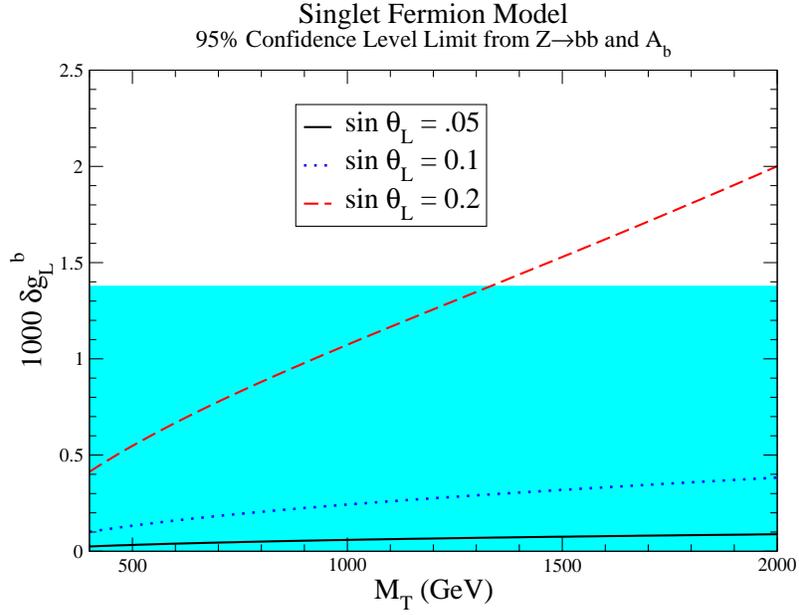} 
\vskip -.14in 
\caption[]{Fermion contributions to $\delta g_L^b$ as a function of
$M_T$ for fixed $\sin\theta_L$ in the top partner singlet model. The $95\%$
confidence level allowed region is shaded in light blue. }
\label{dg_sin}
\end{center}
\end{figure}

The mixing in the charge 2/3 sector also affects $R_b$. In the limit $m_b\rightarrow 0$ and
neglecting the small correlations between $R_b$ and the oblique parameters, only 
$\delta g_L^b$ is affected by the singlet top partner. Its contribution to $\delta g_L^b$ can be found from the 
general analysis of Ref.~\cite{Bamert:1996ci},
\begin{equation}
\delta g_L^b=
{g^2\over 64 \pi^2} 
s_L^2\biggl(
f_1(x,x^\prime)+c_L^2
f_2(x,x^\prime)\biggr)\, ,
\end{equation}
where $x=m_t^2/M_W^2$ and $x^\prime=M_T^2/M_W^2$. 
The Standard Model top contribution has been subtracted following the definition of Eq.~\ref{zdef}. 
In the heavy mass limit, $x,x^\prime>>1$,
\begin{eqnarray}
f_1(x,x^\prime)&=&
x^\prime -x+3\log\biggl({x^\prime\over x}\biggr)
\nonumber \\
f_2(x,x^\prime)&=&
-x-x^\prime +{2 x x^\prime\over x^\prime-x}
\log\biggl({x^\prime \over x}\biggr)
\, .
\label{f1def}
\end{eqnarray}
The contribution to $\delta g_L^b$ from top singlet partners is shown in Fig.~\ref{dg_sin} as
a function of $M_T$ for fixed $\sin\theta_L$, along with the $95\%$ confidence
level region allowed from the fit of Section~\ref{bfits}. We use the exact one-loop
calculation of $\delta g_L^b$ in the top singlet partner model, following 
Refs.~\cite{Bamert:1996ci,Anastasiou:2009rv}.
The relatively large contributions from $R_b$ can be
understood by looking at the leading terms for $m_t, M_T >> M_W$,
\begin{equation}
\delta {g}_L^b=
\delta {g}_L^{b,SM}	s_L^2
\biggl[-(1+c_L^2)+s_L^2 r 
+ 2 c_L^2 {r \over r - 1} \log(r)\biggr] \, .
\end{equation}
Again we see that the heavy top partner decouples only 
when the parameters in the
mass matrix are such that $s_L\sim {v\over M_T}$. 
Furthermore, its contributions to $\delta {g}_L^b$ and to the $T$ parameter 
(Eq.~\ref{singapprox}) are both positive and strongly correlated.
A large, positive contribution to  $\delta {g}_L^b$  from the singlet also 
implies a large contribution to the $T$ parameter. 
This correlation was already pointed out in the context of composite Higgs models in 
Refs.~\cite{Carena:2006bn,Carena:2007ua,Barbieri:2007bh}. 

\begin{figure}]
\begin{center}
\includegraphics[bb=14 38 509 390,scale=0.6]{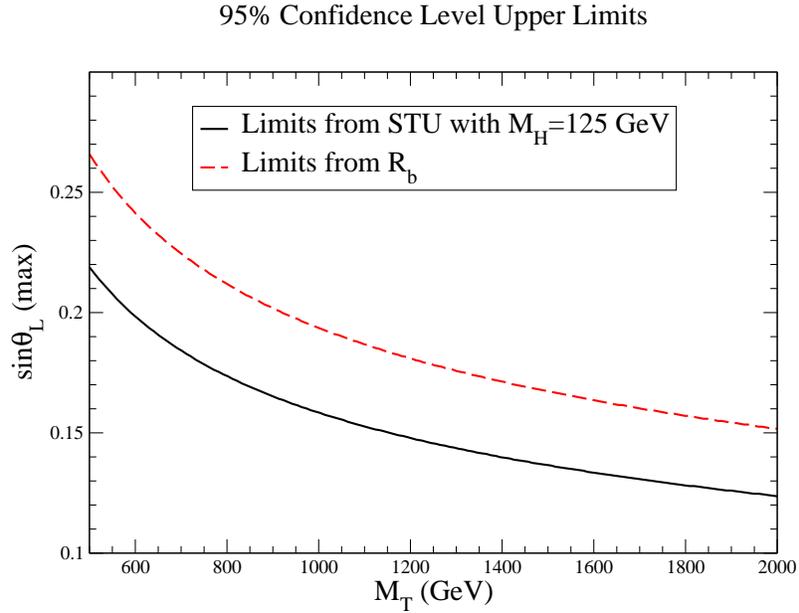} 
\vskip -.14in 
\caption[]{Maximum allowed $\sin\theta_L$ in the top partner singlet model
from oblique measurements (black solid) and $R_b$ (red dashed).}
\label{fg:dg_stu}
\end{center}
\end{figure}
Combining the new physics contribution with the Standard Model top quark contribution, 
\begin{eqnarray}
\delta {\tilde {g}}_L^b 
&=&
\delta g_L^{b,SM}+\delta g_L^b 
=
\delta g_L^{b,SM}
\biggl[
c_L^4 + s_L^4 r 
+2 s_L^2 c_L^2
{r \over r-1} \log(r) \biggr]\, ,
\end{eqnarray}
and the net effect of the top partner is to increase ${ {\delta {\tilde{g}}}}_L^b$
and hence reduce $R_b$.

A comparison of the maximum value of $\sin\theta_L$ allowed by the
fit to $R_b$ and $A_b$ and
by the experimental limits arising from the fit to 
$S$, $T$ and $U$ (with $M_H=125$ GeV) is shown in Fig.~\ref{fg:dg_stu}, where it is apparent that
the most stringent restrictions in the top partner singlet model come from the oblique parameters.

\section{Top Partner Doublet}

\subsection{Model with Top Partner Doublet}

In this section, we consider a model which has in addition to the
Standard Model fields a vector $SU(2)_L$ doublet~\cite{Lavoura:1992np, Cacciapaglia:2010vn},
\begin{equation}
\psi^2_L=
\left( 
\begin{matrix} 
{\cal {T}}_L^2\\
{\cal {B}}_L^2
\end{matrix}\right),
\quad
\psi^2_R=
\left(
\begin{matrix}
{\cal {T}}^2_R\\
{\cal {B}}^2_R\end{matrix}
\right)\, .
\end{equation}
The most general fermion mass terms allowed in the Lagrangian are 
\begin{eqnarray}
-{\cal L}_{M,2}&=&-{\cal L}_M^{SM}+
\lambda_6
{\overline {\psi}}^2_L
H
{\cal {B}}^1_R+
\lambda_7
{\overline {\psi}}^2_L
{\tilde {H}}
{\cal {T}}^1_R+
\lambda_8
{\overline {\psi}}^2_L
\psi^2_R+
\lambda_9
{\overline {\psi}}^1_L
\psi^2_R + {\rm h.c.}
\nonumber \\
&=&{\overline {\chi}}_L^t
\biggl[U_L^{t}
M^t_{(2)} U_R^{t \dagger}\biggr]\chi_R^t +
{\overline {\chi}}_L^b
\biggl[U_L^{b}
M^b_{(2)}U_R^{b \dagger}\biggr]\chi_R^b +{\rm h.c.}
\, ,
\end{eqnarray}
where $\chi_{L,R}^t$ are given by Eq.~\ref{chit} and
\begin{equation}
\chi_{L,R}^b\equiv
\left( \begin{matrix} 
b_{L,R}\\
B_{L,R}\end{matrix}\right)
\equiv U_{L,R}^b
\left(\begin{matrix}
{\cal{B}}^1_{L,R}\\{\cal{B}}^2_{L,R}
\end{matrix}
\right)\, .
\end{equation}

We can always rotate $\psi_2$ such that $\lambda_9=0$. So without loss
of generality 
\begin{equation}
M^t_{(2)}=\left(
\begin{matrix}\lambda_2{v\over\sqrt{2}}&0\\
\lambda_7{v\over\sqrt{2}}&\lambda_8\end{matrix}
\right),\quad
M^b_{(2)}=\left(
\begin{matrix}\lambda_1{v\over\sqrt{2}}&0\\
\lambda_6{v\over\sqrt{2}}&\lambda_8\end{matrix}
\right)\, .
\end{equation}
Because of the $SU(2)$ symmetry 
\begin{eqnarray}
M_{(2),22}^t & = & M_{(2),22}^b \, .
\label{angrel}
\end{eqnarray}
Diagonalizing the mass matrices now requires four unitary mixing matrices,
$U_L^t, U_R^t, U_L^b,U_R^b$,
\begin{eqnarray}
U_L^t&=&
\left(\begin{matrix}
\cos\theta_L^t& -\sin\theta_L^t\\
\sin\theta_L^t& \cos\theta_L^t\end{matrix}
\right),\quad 
U_R^t=
\left(\begin{matrix}
\cos\theta_R^t & -\sin\theta_R^t\\
\sin\theta_R^t & \cos\theta_R^t\end{matrix}
\right) , \nonumber \\
U_L^b&=&
\left(\begin{matrix}
\cos\theta_L^b& -\sin\theta_L^b\\
\sin\theta_L^b& \cos\theta_L^b\end{matrix}
\right),\quad 
U_R^b=
\left(\begin{matrix}
\cos\theta_R^b & -\sin\theta_R^b\\
\sin\theta_R^b & \cos\theta_R^b\end{matrix}
\right) \, .
\end{eqnarray}

There are five input parameters in the Lagrangian. We will take the five 
independent physical
parameters to be the physical masses, $m_t,M_T$ and $m_b,M_B$ (with $m_t$
and $m_b$ being the mass of the Standard Model-like fermions) and the right-handed mixing
angle in the charge -1/3 sector, $\theta_R^b$. It is straightforward 
to find relationships among the mixing angles: 
\begin{eqnarray}
(\sin\theta_R^t)^2&=& {(\sin\theta_L^t)^2\over (\sin\theta_L^t)^2+(\cos\theta_L^t)^2{m_t^2\over M_T^2}} 
\nonumber \\
(\sin\theta_R^b)^2&=& {(\sin\theta_L^b)^2\over (\sin\theta_L^b)^2+(\cos\theta_L^b)^2{m_b^2\over M_B^2}} 
\nonumber \\
(\sin\theta_R^b)^2(m_b^2-M_B^2) +M_B^2&=& 
(\sin\theta_R^t)^2(m_t^2-M_T^2)+M_T^2\, .
\label{doubangs}
\end{eqnarray}
For small mixing, the left-handed angles are always suppressed by the heavy mass scale relative to the right-handed 
angles of the same charge sector,
 \begin{equation}
 \sin^2\theta_L^{t,b}\sim {m_{t,b}^2\over M_{T,B}^2}\sin^2\theta_R^{t,b}\, .
 \end{equation} 
If the mass splitting between $B$ and $T$, $\delta\equiv M_T-M_B$, is small, 
${\mid\delta \mid\over M_T}<<1 $,
\begin{equation}
 (\sin\theta_R^t)^2=(\sin\theta_R^b)^2
 +(\cos\theta_R^b)^2{2\delta\over M_T}+
 {\cal O}\biggl({\delta^2\over M_T^2},{m_t^2\over M_T^2},{m_b^2\over M_T^2}\biggr)
 \, .
\end{equation} 

The charged current interactions are 
\begin{eqnarray}
{\cal L}^{CC}_2&=&{g\over \sqrt{2}} W^{\mu +} \biggl\{\biggl[\Sigma_{i=1}^2
{\overline \psi}^i_L\gamma_\mu \sigma^-
\psi_L^i \biggr]+
{\overline \psi}^2_R\gamma_\mu
\sigma^-\psi_R^2 \biggr\} +{\rm h.c.}
\nonumber \\
&=& {g\over \sqrt{2}} W^{\mu +}  \biggl\{
{\overline \chi}^t_L\gamma_\mu U_L^tU_L^{b,\dagger}
\chi_L^b +
{\overline \chi}^t_R\gamma_\mu
U_R^tU_R^{b,\dagger}\chi_R^b \biggr\}+ {\rm h.c.} \, ,
\end{eqnarray}
where
\begin{equation}
\sigma^-
=\left(
\begin{matrix}
0&1\\0&0
\end{matrix}\right)\, .
\end{equation}

The neutral current interactions are given by Eq.~\ref{neutc}, with 
\begin{eqnarray}
\delta g_L^i&=&\delta g_L^{ij}=0 \; ,\qquad \quad i,j={\hbox{$t,T,b,B$}}\nonumber \\
\delta g_R^t&=&T_3^t\sin^2\theta_R^t \; , \quad\quad 
\delta g_R^T=T_3^T\cos^2\theta_R^t \;  , \quad\quad 
\delta g_R^{tT}=-{1\over 2}\sin \theta_R^t\cos\theta_R^t\nonumber \\
\delta g_R^b&=&T_3^b\sin^2\theta_R^b \; , \quad\quad 
\delta g_R^B=T_3^b\cos^2\theta_R^b \; , \quad\quad 
\delta g_R^{bB}=\phantom{-}{1\over 2}\sin \theta_R^b\cos\theta_R^b\, .
\end{eqnarray}
In the doublet top-partner model, only the right-handed Standard Model-like singlets 
${\cal T}_1^R$, ${\cal B}_1^R$ have Yukawa type mixing with the new quarks, and 
therefore only the interactions in the right-handed sector are modified.
Finally, the Higgs couplings are given by 
\begin{eqnarray}
{\cal L}^h_2 & =& 
-c_{tt}{m_t\over v} {\overline t}_Lt_Rh \,
- \, c_{TT}{M_T\over v} {\overline T}_LT_Rh \,
- \, c_{tT} {m_t\over v} {\overline t}_L T_Rh\,
- \, c_{Tt} {M_T\over v} {\overline T}_Lt_Rh \, +
\nonumber \\
&&
-c_{bb}{m_b\over v} {\overline b}_Lb_Rh
-c_{BB}{M_B\over v}{\overline B}_LB_Rh
-c_{bB}{m_b\over v} {\overline b}_LB_Rh
-c_{Bb}{M_B\over v} {\overline B}_Lb_Rh +{\rm h.c.}\, ,
\end{eqnarray}
where
\begin{eqnarray}
c_{tt}&=&\cos^2 \theta_R^t \, , \qquad 
c_{TT}=\sin^2 \theta_R^t\, , \qquad \,
c_{tT}=c_{Tt}=\sin \theta_R^t\cos \theta_R^t
\nonumber \\
c_{bb}&=&\cos^2 \theta_R^b \, , \qquad  c_{BB}=\sin^2 \theta_R^b\, , \qquad c_{bB}=c_{Bb}=\sin \theta_R^b\cos \theta_R^b\, .
\label{yukdoub}
\end{eqnarray}
The derivation of these couplings follows the lines of Eq.~\ref{eq:why_UR_vanishes_singlet}, 
just with
\begin{equation}
		v H^t_{(2),ks} = M^t_{(1),ks} \delta_{s 1} \, .	
\end{equation}

\subsection{Experimental Limits on Doublet Fermion Model}
\begin{figure}
\begin{center}
\includegraphics[bb=14 38 509 390,scale=0.6]{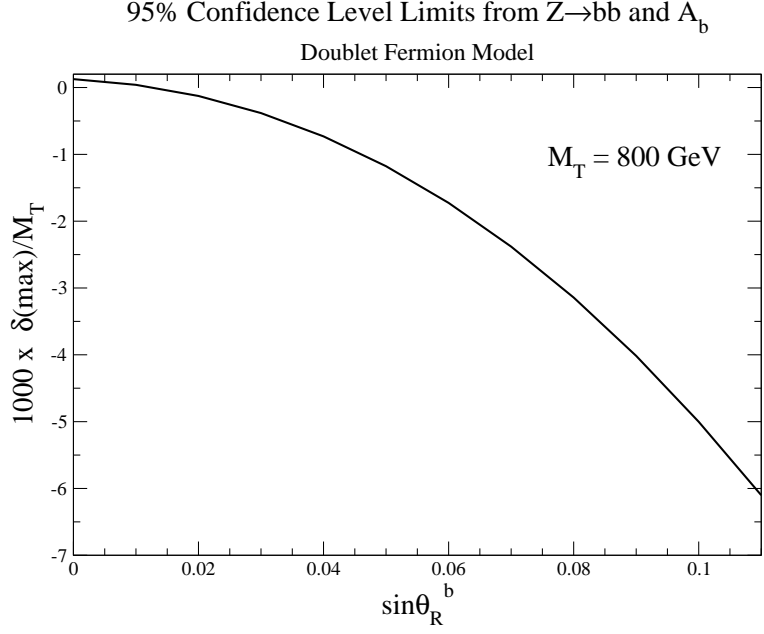} 
\vskip .25in 
\caption[]{
Maximum value of ${\delta\over M_T}$ allowed at the $95\%$ confidence level 
from $R_b$ and $A_b$ 
as a function of $\sin\theta_R^b$ in the doublet fermion model. }
\label{fg:zbb_doublet}
\end{center}
\end{figure}

The decay $Z\rightarrow b{\overline b}$ puts a strong restriction on $\sin \theta_b^R$
since mixing in the right-handed $b$-quark sector 
contributes to $\delta g_R^b$ at tree level,
\begin{equation}
\delta g_R^b=-{1\over 2}(\sin\theta_R^b)^2\, .
\end{equation}
From Eq.~\ref{grfit}, the mixing angle in the right-handed $b$ sector is 
highly constrained,
\begin{equation}
| \sin\theta_R^b | < 0.115 \, . 
\end{equation}
The contribution to $\delta g_L^b$ in the left-handed sector occurs at one-loop.
Subtracting
out the Standard Model contribution, in the limit $x,x^\prime >> 1$, the approximate
result in the doublet fermion model is~\cite{Bamert:1996px}
\begin{equation}
\delta g_L^b={g^2\over 64 \pi^2}
\biggl\{
 \sin^2(\theta_L^t-\theta_L^b)
 f_1(x,x^\prime)
+(\sin\theta_R^t)^2
\cos^2(\theta_L^t-\theta_L^b)f_3(x,x^\prime)
\biggr\}\, ,
\end{equation}
where $f_1(x,x^\prime)$ is defined in Eq.~\ref{f1def} and
\begin{eqnarray}
f_3(x,x^\prime)&=&
-x+{3\over 2}\biggl(1+{x\over x^\prime}\biggr)
+\biggl({x^\prime +x\over 2}-3\biggr)
{x\over (x^\prime -x)}
\log\biggl({x^\prime\over x}\biggr) \, .
\end{eqnarray}
In Fig.~\ref{fg:zbb_doublet},
 we scan over $\sin\theta_R^b$ and ${\delta\over M_T}$ and
use the relationships of Eq.~\ref{doubangs}
 to find the remaining parameters. We use the exact
one-loop result for $\delta g_b^L$ following Refs.~\cite{Bamert:1996px,Anastasiou:2009rv}, and determine 
the $95\%$ confidence level upper bound on ${\delta\over M_T}$. 
Because of the tree level mixing in the $b$ sector
in this model, along with the
relationships of Eq.~\ref{doubangs},
the heavy fermions are required to be approximately degenerate, as is clear from 
Fig.~\ref{fg:zbb_doublet}. This result is relatively insensitive to $M_T$.

\begin{figure}
\begin{center}
\includegraphics[bb=14 38 509 390,scale=0.6]{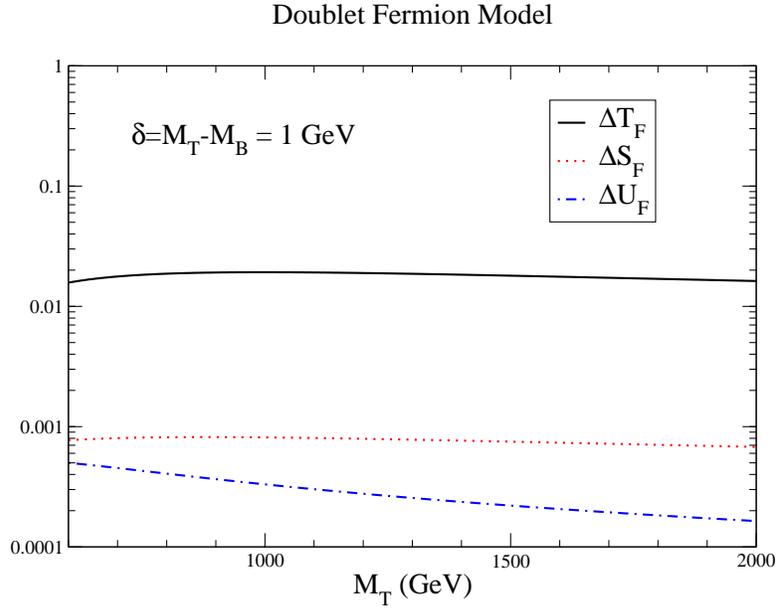} 
\vskip .25in 
\caption[]{
Oblique parameters in the doublet fermion model in the limit $\theta_R^b=0$ and $m_b \to 0$
as a function of $M_T $. We fix $\delta=M_T-M_B=1$~GeV.
}
\label{fg:db_stu}
\end{center}
\end{figure}

Since $\sin\theta_R^b$ is constrained to be quite small,
we will consider the oblique
parameters in the limit $\theta_R^b \!\!=\!0$. From
Eq.~\ref{doubangs}, $\theta_R^b\!\!=\!0$
implies $\theta_L^b\!\!=\!0$ and the free parameters are $m_t, m_b$, $M_T$ and $M_B$. 
Our results will always be expressed in terms of $\delta\equiv M_T-M_B$.
In the $\theta_R^b=0, m_b \to 0$ limit, the oblique parameters are well 
approximated by~\cite{Lavoura:1992np,Maekawa:1995ha} 
\begin{eqnarray}
\Delta T_F(
{\hbox{approx}})
&=&
4 \, T_{SM} {\delta\over M_T}\biggl[2\log(r)-3 + {5 \log(r)-3 \over r} \biggr] 
\nonumber \\
\Delta S_F({\hbox{approx}})&=&{N_C\over 9 \pi }{\delta\over M_T}\biggl[
4 \log(r)-7 + {4 \log(r) + 7 \over r}
\biggr]
\nonumber \\
\Delta U_F({\hbox{approx}})&=&{N_C\over 9\pi}{\delta\over M_T}
\biggl[3+{6 \log(r) - 17 \over r}\biggr]\, ,
\end{eqnarray}
and
\begin{equation}
\delta g_b^{L} = \delta g_b^{L,SM}  {\delta\over M_T}\biggl[\log(r)-4 + 3 \, {\log(r)-2 \over r} \biggr] \,. 
\end{equation}
It is apparent that decoupling of the effects of the new fermions
occurs in the isospin conserving limit, ${\delta\over M_T}\rightarrow 0$. 
Note also that $\delta g_b^{L}$ changes sign for $M_T \simeq 7 m_t$. 
The oblique parameters are shown in Fig.~\ref{fg:db_stu} for $\delta=1$~GeV. As in the singlet model,
$\Delta T_F>>\Delta S_F, \Delta U_F$. 

\begin{figure}
\begin{center}
\includegraphics[bb=14 38 509 390,scale=0.6]{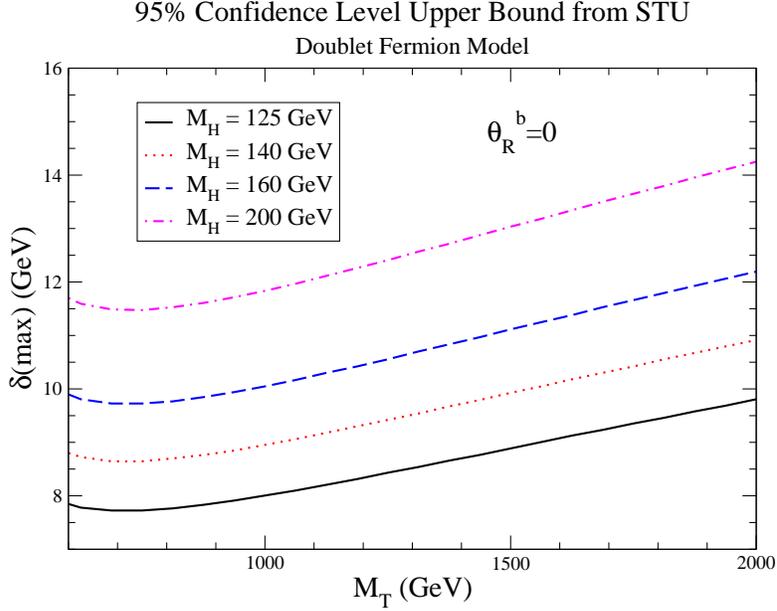} 
\vskip .25in 
\caption[]{Maximum mass splitting $\delta=M_T-M_B$ as a function of $M_T $
allowed at the $95\%$ confidence level from the oblique parameters in the doublet fermion 
model for $\theta_R^b=0$
.}
\label{fg:db_fit}
\end{center}
\end{figure}
The limits coming from the oblique parameters are found from a global fit to $S$, $T$ and $U$ as described
in Section~\ref{stufits}. For $\theta_R^b=0$, the $95\%$ upper limit on the mass splitting, $\delta$, is shown 
in Fig.~\ref{fg:db_fit} as a function of $M_T$. For $M_H=125$~GeV and $M_T\sim 1$~TeV, 
the experimental constraints on the oblique parameters require $\delta \lesssim 8$~ GeV.
As shown in Fig.~\ref{fg:db_fit}, it is possible to compensate for 
the negative contribution to $T$ from 
a heavier Higgs boson by a larger mass splitting $\delta$, which generates
a positive contribution to $\Delta T_F$.
However, the limits from $A_b$ and $R_b$ in this model are much more stringent than those coming
from the oblique parameters.

\section{Phenomenology}
The new fermions affect the gluon fusion production rate,
which at lowest order is given by~\cite{Wilczek:1977zn, Ellis:1979jy, Georgi:1977gs} 
\begin{equation}
\sigma(gg\rightarrow H)=
{\alpha_s^2\over 1024 \pi v^2}
\mid\Sigma_qc_{qq}F_{1/2}(\tau_q)\mid^2
\delta\biggl(1-{{\hat s}\over M_H^2}\biggr)\, .
\end{equation}
The sum is over $t,b,T$ in the singlet fermion model and over $t,b,T,B$ in the doublet 
model, the Yukawa couplings normalized to the Standard Model values $c_{qq}$ are given 
in Eqs.~\ref{yuksing} and~\ref{yukdoub}, and
\begin{eqnarray}
\tau_q \, &=&{4M_q^2\over M_H^2}\nonumber\\
F_{1/2}&=&-2\tau_q\biggl[
1+(1-\tau_q)f(\tau_q)\biggr]\nonumber\\
f(\tau_q)&=&\left\{
\begin{matrix}
 [\sin^{-1}\biggl(\sqrt{{1\over\tau_q}}\biggr)\biggr]^2,& {\hbox {if}}~ \tau_q\ge 1\\
-{1\over 4}\biggl[\log\biggl({1+\sqrt{1-\tau_q}\over 1-\sqrt{1-\tau_q}}-i\pi\biggr]^2,
&{\hbox {if}} ~\tau_q<1\, .
\end{matrix}\right.
\end{eqnarray}

We compute the gluon fusion production cross section through NNLO 
using the program 
{\texttt iHixs}~\cite{Anastasiou:2011pi}. 
{\texttt iHixs} allows the calculation of the cross section at NNLO 
for extensions of the Standard Model with an arbitrary number of heavy quarks having 
non-standard Yukawa interactions~\cite{Furlan:2011uq}, and puts the predictions on
a firm theoretical basis. 
At NNLO, there are contributions which mix quark loops of
different flavors (e.g. $t$ and $T$) and cannot be obtained by a simple rescaling of
the lowest order cross section.
We scan the parameter space allowed by the precision electroweak data as determined in the
previous sections and discuss the maximum deviations from the Standard Model predictions.

\subsection{Top Partner Singlet Model}
\begin{figure}
\begin{center}
\includegraphics[bb=14 38 509 390,scale=0.6]{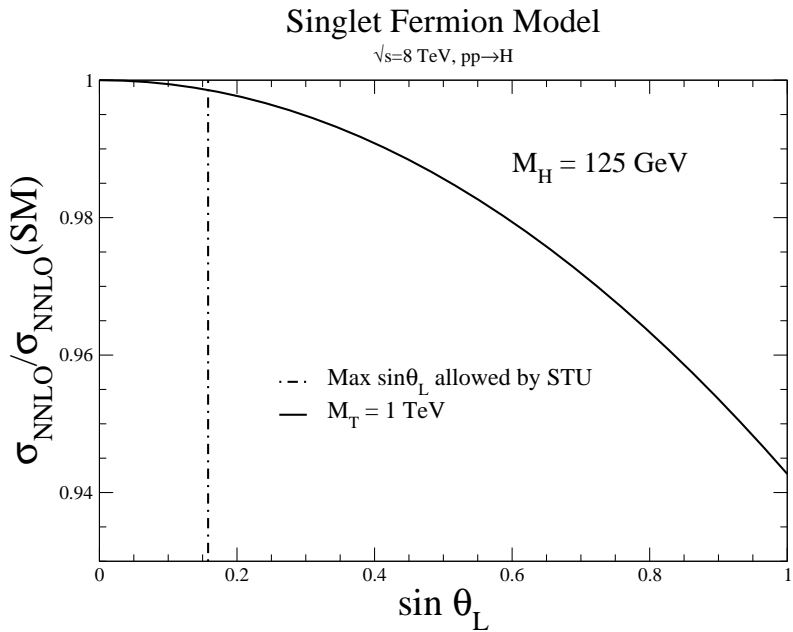} 
\vskip .25in 
\caption[]{The ratio of the NNLO Higgs cross section in the top partner singlet model
normalized to the Standard Model prediction as a function
of $\sin\theta_L$ for $M_H=125$~GeV and $\sqrt{s}=8$~TeV. The vertical line 
represents the maximum value of $\sin\theta_L$ allowed by electroweak precision
measurements.
}
\label{fg:cross1}
\end{center}
\end{figure}
The deviation from the Standard Model prediction 
of the NNLO Higgs production cross section as a function of the mixing angle 
in the top partner singlet model is 
shown in Fig.~\ref{fg:cross1}.
The largest value of $\sin\theta_L$ allowed
by the precision electroweak limits derived in Section~\ref{sec:exp_bounds_singlet} is also shown. 
As $\sin\theta_L$ increases, the mixing with the Standard Model-like top quark
becomes significant, causing a suppression of the rate. 
This can be understood
from the heavy mass limit ($ m_t,M_T >> {M_H \over 2}$) of the lowest
order cross section, where the gluon fusion
rate scales as
\begin{equation}
{\sigma_{\rm{Singlet}} \over \sigma_{SM}}
\sim
1-{7\over 60}{M_H^2\over m_t^2} s_L^2 \biggl(1-{m_t^2\over M_T^2}\biggr)\, .
\end{equation}
However, only the region to the left
of the dot-dash line in Fig.~\ref{fg:cross1} is allowed by the precision electroweak
measurements, making the Higgs boson production
rate in this model almost identical to the Standard Model rate.
In contrast with composite~\cite{Falkowski:2007hz} or Little 
Higgs~\cite{Low:2009di, Berger:2012ec} models, which typically have a sizeable reduction of
the Higgs production cross section relative to the Standard Model, 
in models with vector fermions the suppression is negligible because of the decoupling properties
discussed in the previous sections.
The uncertainty on the Standard Model cross section coming from scale, PDF,
and $\alpha_s$ uncertainties is roughly $15-20\%$~\cite{LHCHiggsCrossSectionWorkingGroup:2011ti}, 
so the extremely small deviation from
the Standard Model prediction in the top partner singlet model is unobservable. 
\begin{figure}
\begin{center}
\includegraphics[bb=14 38 509 390,scale=0.6]{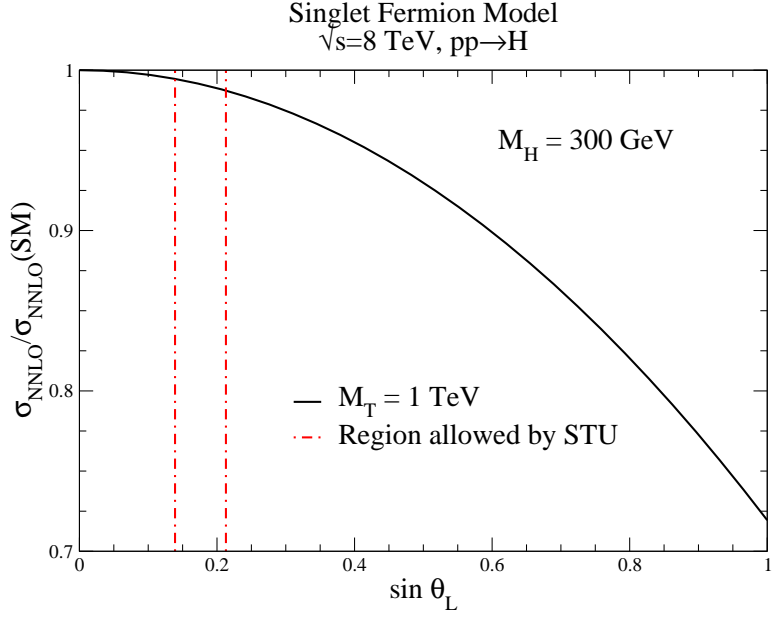} 
\vskip .25in 
\caption[]{The ratio of the NNLO Higgs cross section in the top partner singlet model
normalized to the Standard Model prediction as a function
of $\sin\theta_L$ for $M_H=300$~GeV and $\sqrt{s}=8$~TeV. The region
between the vertical lines 
represents the region of $\sin\theta_L$ allowed by electroweak precision
measurements.
}
\label{fg:cross2}
\end{center}
\end{figure}
\begin{figure}
\begin{center}
\includegraphics[bb=14 38 509 390,scale=0.6]{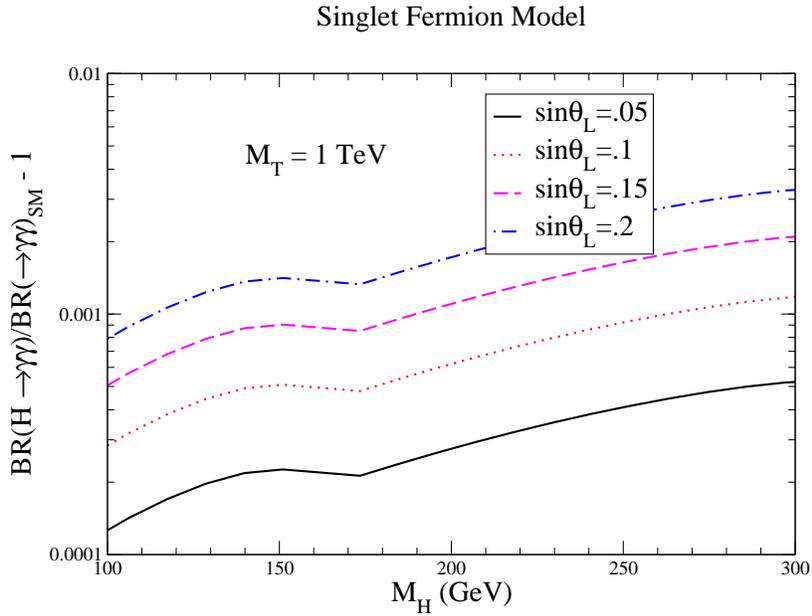} 
\vskip .25in 
\caption[]{Deviation of the $H\rightarrow \gamma \gamma$ branching ratio in the
top partner singlet model from the Standard Model prediction.
}
\label{fg:decay}
\end{center}
\end{figure}
The cross section for a heavier Higgs boson of mass $M_H=300$~GeV is
shown in Fig.~\ref{fg:cross2}. In this case, there is a region of mixing angles, $\sin \theta_L$, which is allowed
by the precision electroweak measurements. Again, there is a slight, but unobservable, suppression of the NNLO
rate relative to the Standard Model rate.

The loop-mediated Higgs decays to $\gamma \gamma$, $Z\gamma$ and $gg$ are also affected by the presence
of top fermion partners. Fig.~\ref{fg:decay} shows the deviation of the branching ratio to $\gamma \gamma$
from the Standard Model prediction. For small mixing, this deviation is always less than one percent.

\subsection{Top Partner Doublet Model}
\begin{figure}[b]
\begin{center}
\vskip .35in 
\includegraphics[bb=14 38 509 390,scale=0.6]{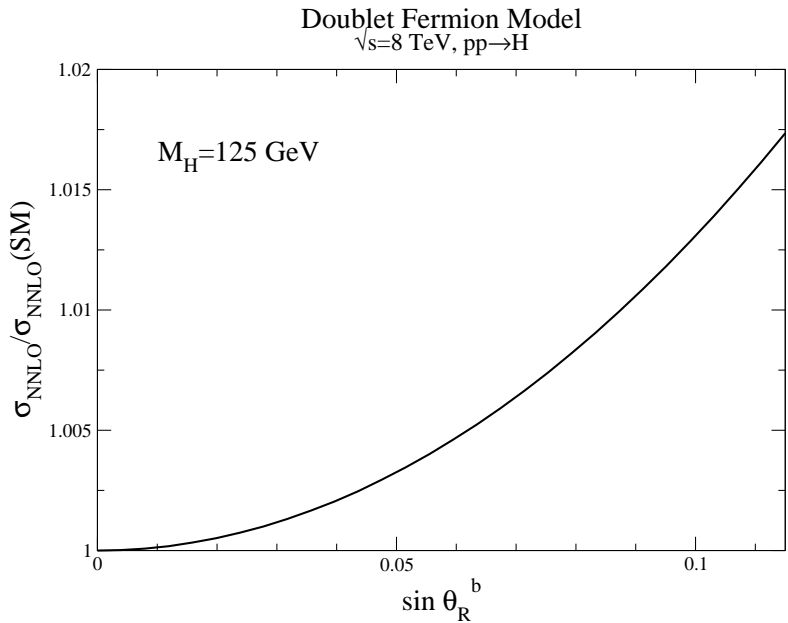} 
\vskip .15in 
\caption[]{The ratio of the NNLO Higgs cross section in the top partner doublet model
normalized to the Standard Model prediction as a function
of $\sin\theta_R^b$ for $M_H=125$~GeV, $\sqrt{s}=8$~TeV,
and $M_T=M_B=1$~TeV.
}
\label{fg:cross3}
\end{center}
\end{figure}
The deviation from the Standard Model prediction 
for the NNLO gluon fusion cross section for Higgs production in the top partner doublet model
(computed using {\texttt iHixs}) is shown in Fig.~\ref{fg:cross3}. Also in this 
case the maximum difference from the Standard Model in the allowed regions of parameter 
space (Fig.~\ref{fg:zbb_doublet}) is always less than a few percent. 
This result can be understood by considering
the heavy mass limit of the lowest order cross section for the gluon
fusion production of the Higgs,
\begin{eqnarray}
{\sigma_{\textrm{Doublet}}\over \sigma_{SM}}&\simeq &
\biggl(1+\sin^2\theta_R^b\biggr)
\biggl[1+\sin^2\theta_R^b-
{7\over 60} {M_H^2\over m_t^2}
\biggl( {2 r-1 \over r} \sin^2\theta_R^b
\nonumber \\
&&
+ {2 \delta \over M_T} - {2 \delta \over M_T}  {r+1 \over r}  \sin^2\theta_R^b
\biggr)\biggr] \,.
\end{eqnarray}
From the fits to $A_b$ and $R_b$, the maximum value of $ \sin\theta_R^b$
is restricted to be 0.115, which implies 
\begin{eqnarray}
{\sigma_{\textrm{Doublet}} \over \sigma_{SM}}& \lesssim &
 \biggl(1+ \sin^2\theta_R^b\biggr)^2 \simeq 1.03\, .
\end{eqnarray}

Similarly, the deviations from the Standard Model in the Higgs decays to $\gamma \gamma$, $Z\gamma$ and $gg$ 
and in $H \to b \bar{b}$, which is affected at tree level, are not observable due to the small mixings 
and mass splitting allowed.

\section{Conclusions}
We have considered the effects on the gluon fusion Higgs boson production at NNLO
from heavy vector quarks of  charge 2/3
and -1/3. Since the new quarks are vector-like, their
couplings to the Higgs boson are suppressed by mixing angles relative to the Standard Model
Yukawa couplings. These mixing angles are restricted to be small by precision electroweak measurements. 
The most stringent bounds come from the oblique parameters for a vector singlet top-partner, and from 
$A_b$ and $R_b$ for an extension of the Standard Model with an additional vector doublet. 
Because of the small mixing angles allowed, in these models the Higgs boson production rate as well as its decay branching 
ratios are essentially those of the Standard Model. The
scenarios we have presented will be extremely difficult to disentangle from the Standard Model
without the observation of direct production of the heavy fermions.  Vector doublet fermions with a non-standard
hypercharge assignment are less restricted by precision electroweak measurements~\cite{Cacciapaglia:2010vn} and the mixing
angles between the $t-b$ sector and the new fermion sector can be larger than in the cases we considered.
  However, even in this case, the low
energy theorems for Higgs production require that the Higgs cross section approach the Standard Model result
for large fermion masses.  

If a Higgs boson is found at the LHC, attention will turn to understanding its properties.  By
performing global fits to the measured rates, information can be gleaned from the various cross section
times branching ratio channels.  For a light Higgs boson, it is likely that the dominant production channel
will be gluon fusion, even in models with new physics. 
In this case, the rates are sensitive not only to a rescaling of the Standard Model couplings, but also
to the effects of new particles which couple to the Higgs boson and contribute to the decay rates.
Numerous preliminary attempts have been
made to use current LHC data to discern differences from the Standard
Model~\cite{ Low:2009di, Dobrescu:2011aa,Englert:2011aa,Espinosa:2012qj,Carmi:2012yp,Azatov:2012bz,
Espinosa:2012ir,Giardino:2012ww,Ellis:2012rx,Klute:2012pu}.
Our scenario with vector fermions demonstrates the difficulty of these indirect determinations
of new physics -- it is (un)fortunately not difficult to construct models which give Higgs signals 
indistinguishable from the Standard Model. 

\section*{Acknowledgements}
We would like to thank G. Panico for useful discussions. This work is supported by the United States Department of Energy under
Grant DE-AC02-98CH10886.

\section*{Appendix: Two Point Function for Arbitrary Fermion Coupling}
The contributions to the gauge boson two point functions from fermion loops parametrized by the interaction 
\begin{equation}
{{\cal L}=\overline f}_1 
\biggl(C_{LX}^{f_1f_2} P_L
+C_{RX}^{f_1f_2} P_R\biggr)\gamma_\mu f_2 V^\mu\,, 
\end{equation}
for $V=W,Z,\gamma$ are~\cite{Chen:2003fm,Jegerlehner:1991dq}
\begin{eqnarray}
\Pi_{XY}&=&
-{N_c\over 16 \pi^2}\biggl\{
{2\over 3}\biggl(C_{LX}^{f_1f_2}C_{LY}^{f_1f_2}+
C_{RX}^{f_1f_2}C_{RY}^{f_1f_2}\biggr)
\biggl[m_1^2+m_2^2-{p^2\over 3}-
\biggl(A_0(m_1)+A_0(m_2)\biggr)
\nonumber \\
&&+{m_1^2-m_2^2\over 2 p^2}
\biggl(A_0(m_1)-A_0(m_2)\biggr)
+{2p^4-p^2(m_1^2+m_2^2)-(m_1^2-m_2^2)^2\over 2 p^2}
B_0(m_1,m_2,p^2)\biggr]
\nonumber \\
&&
+2m_1m_2\biggl(C_{LX}^{f_1f_2}C_{RY}^{f_1f_2}+
C_{RX}^{f_1f_2}C_{LY}^{f_1f_2}\biggr)B_0(m_1,m_2,p^2)\biggr\}
\end{eqnarray}
where
\begin{eqnarray}
A_0(m)&=&\biggl({4\pi\mu^2\over m^2}\biggr)^\epsilon
\Gamma(1+\epsilon)
\biggl({1\over \epsilon}+1\biggr) m^2
\nonumber \\
B_0(m_1,m_2,p^2)&=&\biggl({4\pi\mu^2
\over m_2^2}\biggr)^\epsilon\Gamma(1+\epsilon)
\left[{1\over \epsilon} -f_1(m_1,m_2,p^2)\right]
\end{eqnarray}
and
\begin{equation}
f_1(m_1,m_2,p^2)=\int_0^1 dx \log\biggl(x+{m_1^2(1-x)-p^2x(1-x)\over m_2^2}
\biggr)\, .
\end{equation}
\

\bibliographystyle{unsrt}
\bibliography{main}

\end{document}